\documentclass{jfm}

\usepackage{graphicx}
\usepackage{amsmath}

\usepackage{siunitx}
\usepackage{silence}
\usepackage{caption, subcaption}
\usepackage{lscape}
\usepackage{xcolor}
\usepackage{soul}
\usepackage[normalem]{ulem}
\usepackage{cancel}

\usepackage{placeins}
\usepackage{layouts}

\usepackage{titlesec}
\titleformat{\subsubsection}[block]{\filcenter}
             {\thesubsubsection}{1em}{}
\titleformat{\subsection}[block]{\filcenter}
             {\thesubsection}{1em}{}

\newcommand{\Ta}{\text{Ta}}
\newcommand{\Nuw}{\text{Nu}_\omega}

\newcommand{\Rei}{\text{Re}_i}
\newcommand{\refsec}[1]{section~\ref{#1}}

\newcommand{\reff}[1]{figure~\ref{#1}}
\newcommand{\refF}[1]{Figure~\ref{#1}}
\newcommand{\refe}[1]{equation~\ref{#1}}
\newcommand{\docname}{manuscript } 

\title{Controlling secondary flow in Taylor--Couette turbulence through spanwise-varying roughness}

\shorttitle{%
Spanwise-varying roughness controls secondary flow in TC flow
}

\author{%
Dennis~Bakhuis$^1$,\ns
Rodrigo~Ezeta$^1$,\ns
Pieter~Berghout$^1$,\ns\break
Pim~A.~Bullee$^{1,2}$,\ns
Dominic~Tai$^1$,\ns
Daniel~Chung$^{3}$,\ns
Roberto~Verzicco$^{4,1}$,\ns\break
Detlef Lohse$^{1,5}$\ns 
Sander~G.~Huisman$^{1}$,\ns and\ns
Chao Sun$^{6,1}$\footnote{Email address for correspondence: chaosun@tsinghua.edu.cn}}
\shortauthor{D. Bakhuis and others}

\affiliation{%
$^1$Physics of Fluids Group, Max Planck UT Center for Complex Fluid Dynamics,\break
MESA+ Institute and J.M. Burgers Centre for Fluid Dynamics,\break
University of Twente, P.O. Box 217, 7500 AE Enschede, The Netherlands\\
\vspace{0.8mm}
$^2$Soft matter, Fluidics and Interfaces, MESA+ Institute for Nanotechnology, \break
University of Twente, P.O. Box 217, 7500 AE Enschede, The Netherlands\\
\vspace{0.8mm}
$^3$Department of Mechanical Engineering, University of Melbourne, Victoria 3010, Australia\\
\vspace{0.8mm}
$^4$Dipartimento di Ingegneria Industriale, University of Rome 'Tor Vergata’,\break Via del Politecnico 1, Rome 00133, Italy\\
\vspace{0.8mm}
$^5$Max Planck Institute for Dynamics and Self-Organization, Am Fa{\ss}berg 17, G\"ottingen, Germany\\
\vspace{0.8mm}
$^6$Center for Combustion Energy, Key Laboratory for Thermal Science and Power Engineering of Ministry of Education, Department of Energy and Power Engineering, Tsinghua University, Beijing, China}
\begin{document}
\maketitle
\begin{abstract}
Highly turbulent Taylor--Couette flow with spanwise-varying roughness is investigated experimentally and numerically {(direct numerical simulations (DNS) with an immersed boundary method (IBM))} to determine the effects of the spacing and axial width $s$ of the {spanwise varying roughness} on the total drag and {on} the flow structures. We apply sandgrain roughness, in the form of alternating {rough and smooth} bands to the inner cylinder. Numerically, the Taylor number is $\mathcal{O}(10^9)$ and {the roughness width} is varied {between $0.47\leq \tilde{s}=s/d \leq 1.23$}, where $d$ is the gap width. Experimentally, we explore $\Ta=\mathcal{O}(10^{12})$ and $0.61\leq \tilde s \leq 3.74$. For both approaches the radius ratio is fixed at $\eta=r_i/r_o = 0.716$, with $r_i$ and $r_o$ the radius of the inner and outer cylinder respectively. We present how {the global transport properties and the local flow structures} depend on the boundary conditions set by {the roughness spacing $\tilde{s}$}. Both numerically and experimentally, we find a maximum in the angular momentum transport
{as function of $\tilde s$. This can be atributed}
to the re-arrangement of the large-scale structures triggered by the presence of the rough {stripes, leading to correspondingly large-scale} turbulent vortices.
\end{abstract}
\section{Introduction}\label{sec:spanwiseintro}
{In nearly all industrial applications and geophysical flows, turbulence is partly or completely wall-bounded. In general, boundaries are not smooth but their surface is rather irregular and rough.
Accordingly, such} flows are extensively studied, {although mainly} under the approximation that the roughness is {\textit{homogeneous}} \citep{Jimenez2004}.
Homogeneously rough surfaces have a characteristic length scale $k$ that is much smaller than the largest wall normal length scale $\delta$.
The effects of the roughness in these flows is believed to be confined to the immediate vicinity of the wall (i.e. the roughness sublayer), whereas in the outer, inertial layer, the flow only experiences the effective shear stress of the surface (Townsend's outer layer similarity, \cite{Townsend1976}).
As such, the focus of many studies is to find functional relationships between the parameters describing the roughness geometry and the skin friction coefficient $C_f$ \citep{Flack2010}.
In practice, however, flows are bounded by rough boundaries that, not only vary on the scale of $k$, but also on a much larger scale $s$, being $s=\mathcal{O}(\delta)$. Whereas these variations can occur either laterally (spanwise) or longitudinally (streamwise), we focus here only on the former.
Such examples are found in shipping (i.e.~the formation of stripes of biofouling on ship hulls \citep{Schultz2007}) and geophysical flows (e.g.~the atmospheric flows over spanwise-varying terrain \citep{Ren2011}).

\indent Hitherto, the research is focused on the effects of spanwise-varying rough surfaces on canonical systems of wall-bounded turbulence research, i.e. pipe- \citep{Koeltzsch2002}, boundary layer- \citep{Anderson2015}, and channel-flow \citep{Chung2018}.
The hallmark of flows over these surfaces is the presence of spanwise wall-normal secondary flows of size  $\mathcal{O}(\delta)$, with mean streamwise vorticity. Examples of studies where this has been observed are the works of \citet{Koeltzsch2002} on the effects of convergent and divergent grooves (reminiscent of shark skin) and the work by \cite{Wang2006} on spanwise-varying riverbeds. We note that earlier research dates back to the works of \citet{Hinze1967,Hinze1973} {in} the field of surface stress variations in duct flows.

\indent Following up on the work of \cite{Koeltzsch2002}, \cite{Nugroho2013} set out to perform a parametric study of the converging-diverging riblets surface in a zero pressure gradient boundary layer (BL). They find a thickening of the BL height above the converging regions, and vice versa above the diverging regions. Furthermore, the energy spectra show an increased energy content of the larger scales. \cite{Barros2014} performed stereo particle image velocimetry (PIV) in the spanwise wall-normal plane for the flow over a turbine blade replica and found spanwise variations of the order $\delta$ in the mean velocity field. Within the same configuration, \citet{Mejia-Alvarez2013} identified regions of low momentum pathways (LMPs) and high momentum pathways (HMPs) in the instantaneous fields. Here, LMPs coincide with regions of enhanced turbulent kinetic energy (TKE) and Reynolds shear stress (RSS), and rather remarkably, these regions do seem to occur at recessed roughness heights.
\cite{Willingham2014} found very similar behaviour of the secondary flows for a much more regular surface geometry.
\cite{vanderWel2015} found that only when $s/\delta \gtrapprox 0.5$, where $s$ is the spacing between streamwise aligned Lego\textsuperscript{\textregistered} blocks, secondary flow formation is observed. However, for $s/\delta \lessapprox 0.5$ the secondary flows are confined to the roughness sublayer. Interestingly, contrary to the findings of \citet{Mejia-Alvarez2013}, they find LMPs on top of their elevated blocks, and HMPs in between the roughness strips.
\cite{Yang2017}, however, found $s/H \gtrapprox 0.2$, with $H$ the channel half height, as the threshold for heterogeneous behaviour of the streamwise aligned pyramid elements. By carefully assessing the terms in the transport equation TKE, \cite{Anderson2015} found that spanwise variations of roughness lead to a local imbalance of production and dissipation of TKE, as already proposed by \cite{Hinze1967}.
Since the secondary flows are driven by a spatial gradient of the RSS, they find that the mean secondary flows are Prandtl's secondary flow of the second kind \citep{Bradshaw1987}.
\cite{Medjnoun2018} observed a breakdown of outer layer similarity in the local profiles of the mean flow, turbulent intensity, and the energy spectra, evidently induced by the presence of the secondary vortices.
Finally, \cite{Chung2018} studied the influence of the spacing of idealized (i.e. no geometric induced disturbances to the flow) regions of low shear stress and high shear stress.
They find that for $s/\delta \lessapprox 0.39$ the notion of outer layer similarity is retained.
Interestingly, for $s/\delta \gtrapprox 6.28$, they find a sign reversal of the isovels (stream velocity contour lines), with respect to the orientation of the secondary flows, that remain upwelling over low shear stress regions.

\indent The aforementioned studies were all carried out in systems that lack two featuresg which are intrinsic to many applications, namely the curvature in the streamwise direction (as in turbine blades), and the presence of strong secondary motions (as in the atmospheric boundary layer).

\indent A canonical system in which these two properties can be {observed simultaneously} is the Taylor--Couette (TC) flow.
TC flow is the flow between two coaxially, independently rotating cylinders. Its geometry is characterized by the inner cylinder radius $r_{i}$, outer cylinder radius $r_o$, and their axial length $L$, described by two dimensionless parameters; the radius ratio $\eta=r_i/r_o$ and the aspect ratio $\Gamma=L/d$, where $d=r_o-r_i$ is the gap width between the cylinders. 
Since TC is a closed system, one can directly relate global and local quantities through exact mathematical relations \citep{Eckhardt2007}. The driving {strength} in TC flow is expressed in dimensionless form by the Taylor number: 
\begin{align}\label{eq:Ta}
    \Ta = \frac 14 \sigma d^2 \frac{(r_i+r_o)^2(\omega_i-\omega_o)^2}{\nu^2},
\end{align}
where $\omega_{i,o}$ are the inner and outer angular velocity of the cylinders, respectively, $\nu$ is the kinematic viscosity of the fluid, and $\sigma=\left(\left(1+\eta\right)/\left(2\sqrt{\eta}\right)\right)^4$ is the so-called geometric Prandtl number, in analogy to the Prandtl number in Rayleigh-B\'{enard} convection \citep{Eckhardt2007}. Alternatively, when the outer cylinder is at rest ($\omega_o=0$), the driving {strength} can also be expressed with a Reynolds number based on the inner scales $\text{Re}_i=r_i \omega_i d / \nu$. This Reynolds number and Ta ($\omega_o=0$), are related by $\text{Re}_i = (8 \eta^2/(1 + \eta)^3) \sqrt{\Ta}$.
In TC flow, the angular velocity flux $J^\omega$ is radially conserved. Here, $J^\omega = r^3 (\langle u_r \omega \rangle_{A,t} - \nu  \frac{\partial}{\partial r} \langle \omega \rangle _{A,t})$, where the brackets $\langle \cdot \rangle _{A,t}$ denote averaging over a cylindrical surface and time. The angular momentum flux for the case of laminar flow is $J_{lam}^\omega = 2\nu r_i^2 r_o^2 (\omega_i-\omega_o) /(r_o^2 - r_i^2)$. In this way the response of the flow is  quantified by the dimensionless Nusselt number ($\Nuw$), which is also directly related to the torque $\mathcal{T}$ that is required to drive the cylinders at constant speed, i.e.
\begin{align}\label{eq:spanwisenu}
\Nuw = \frac{J^\omega}{J^\omega_{lam}}= \frac{\mathcal{T}}{2\pi L \rho J_{lam}^\omega}.
\end{align}
Here, $\rho$ is the density of the working fluid. Alternatively, the torque of the system can be non-dimensionalized to form the friction coefficient $C_f = \mathcal{T}/(\rho L \nu^2 \text{Re}_i^2)$, which is directly related to the Nusselt number:
\begin{align}
\Nuw = C_f \omega_i \left( r_o - r_i \right)^2 \left( r_o^2 - r_i^2\right) / \left(4 \pi \nu r_o^2\right).
\end{align}
The inner friction velocity $u_{\tau,i}$ is also related to the torque by $u_{\tau,i}=\sqrt{\mathcal{T}/(2\pi r_i^2 \rho L)}$, which is used to scale quantities in the inner layer. Lastly, a frictional Reynolds number based on the inner scales can be defined as $\text{Re}_\tau=u_{\tau,i}d/(2\nu)$.

Secondary flows are featured in TC flow, in the form of large scale vortices with a mean streamwise vorticity component, the so-called Turbulent Taylor Vortices (TTV). These structures are reminiscent of laminar Taylor vortices, which transition through a series of instabilities into turbulence once the flow becomes unstable \citep{Taylor1923b}. As noted by \citet{Chouippe2014}, the axial wavelength {$\lambda / d$} of the TTVs, \textit{i.e.} the distance between two rolls, is primarily a function of $\eta$ and $\text{Re}$. When $\text{Re}$ is large enough ($\mathcal{O}(10^6)$), the rolls are observed to persist in the system \citep{Huisman2014}. Here, multiple states for $\eta=0.716$ can be observed in a certain regime of counter-rotating cylinders, namely $a\in[0.17,0.51]$, where $a=-\omega_o/\omega_i$ is their rotation ratio. These multiple states are characterized by a change in the number of rolls present in the system and, as a consequence, in their averaged axial wavelength ($\lambda/d = 1.46$ or $\lambda/d = 1.96$). These states, {with the transition between them being strongly hysteretic, even at $\text{Re}=\mathcal{O}(10^6)$\citep{Huisman2014,vanderveen2016a}}, result in different torques for the same rotation rates, which reflects the importance of the large scale structures (TTV) in transporting angular momentum. At pure inner cylinder rotation however ($a=0$), no multiple states are found and the rolls are observed to be less coherent and stable. Finally, we note that the effect of the curvature of the cylinders is quantified by the radius ratio $\eta$, and it has a tremendous impact on the flow organization as was reported by \citet{Ostilla-Monico2014c,Ostilla-Monico2014}. For a detailed review {on turbulent Taylor-Couette flow} we refer the reader to { \citet{Grossmann2016}}.

{Roughness in a TC geometry has been studied in various ways:} \citet{Cadot1997, vandenBerg2003} used obstacle roughness, in the form of axial riblets, to study the scaling of the angular momentum transport with the driving {strength}.
\citet{Zhu2016} investigated the influence of grooves for large Ta ($\mathcal{O}(10^{10})$), and find that at the tips of the grooves, plumes are preferentially ejected. In a more recent work, \citet{Zhu2018} find that by using a similar configuration of rough walls as \citet{vandenBerg2003}, the scaling {$\Nuw \propto \Ta^{1/2}$ predicted by the so-called asymptotic ultimate regime can be achieved.} They attribute this to a dominance of the pressure drag over the viscous drag on the cylinders. Very recently, \cite{Berghout2018} studied the influence of sandgrain roughness in TC flow, and found similarity of the roughness function with the same type of roughness in pipe flow \cite{Nikuradse1933}.
{None of the TC papers described above reported an influence of the roughness variations in the axial {direction, i.e. the spanwise direction}.}

{In this \docname we will fill this gap and study the effects of spanwise-varying roughness in highly turbulent TC flow with $\text{Ta}$ up to $\mathcal{O}(10^{12})$, for the case of pure inner cylinder rotation $a=0$, where secondary flows are present in the form of TTVs.} In particular, we focus on the effect of spanwise-varying roughness on the TTVs and thus, on the global and local response of the flow. We introduce the roughness through a series of stripes which extend along the entire circumference of the inner cylinder (IC). This gives rise to a spanwise (axial) arrangement of roughness which we characterize with the widths of the {roughness} stripe. We conduct both, experiments and direct numerical simulations (DNS) for various {dimensionless stripe widths} $\tilde{s}=s/d$, i.e. the width of the {roughness} stripe normalized with the gap width. The spacing between the roughness stripes is identical to the stripe width, thus a period consisting of one rough and smooth area has a width of $2\tilde s$.

The structure of the \docname is as follows. In \refsec{sec:methods} we introduce the experimental and numerical methods. In \refsec{sec:resultsec} we show the local response of the flow due to the varying roughness arrangement.  In \refsec{sec:resultglob}, we study its effect on the global quantities. In \refsec{sec:results_velprofiles} we link the global and local observations and {explain} the physical mechanism between the interaction of the rolls and the roughness. We finalize the \docname in \refsec{sec:conclusions} with some conclusions and an outlook to future work.

\section{Methods}
\label{sec:methods}
\subsection{Experimental apparatus with spanwise roughness}
\begin{figure}
\centering
\includegraphics{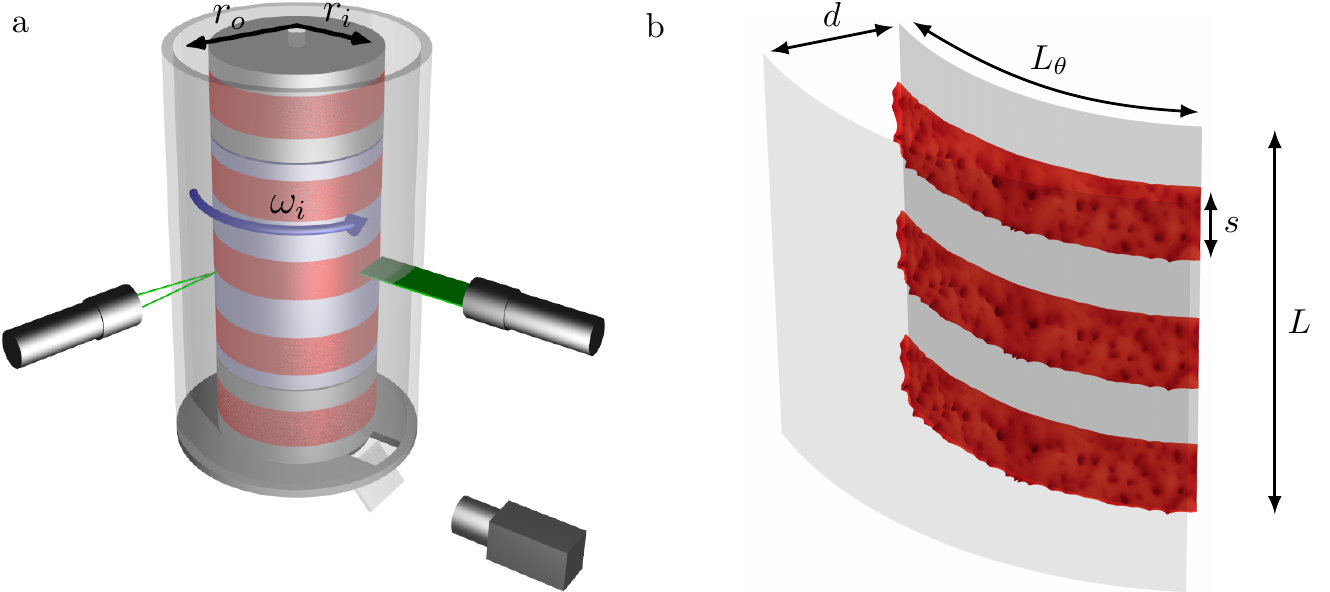}
\caption{a) Schematic of the Twente Turbulent Taylor--Couette {facility} showing the sand paper {roughness} on the inner cylinder in red. PIV measurements in the $r$--$\theta$ plane {become possible thanks to illumination from the side with} a high-power pulsed laser, creating a horizontal sheet. The sheet is imaged through a window in the bottom. Using LDA the azimuthal velocity is measured along the axial direction. The torque is measured in the middle section of the IC, which has a length of $\text{L}_{mid}=\SI{536}{\mm}$. b) Numerical domain for the case of $\tilde s\equiv s/d=0.47$.  {The} sandpaper roughness {is} taken from {a confocal} scan of the material used in the experiment, see \protect\reff{fig:roughness}.}
\label{fig:setupnum}
\end{figure}

\begin{figure}
    \centering
    \includegraphics{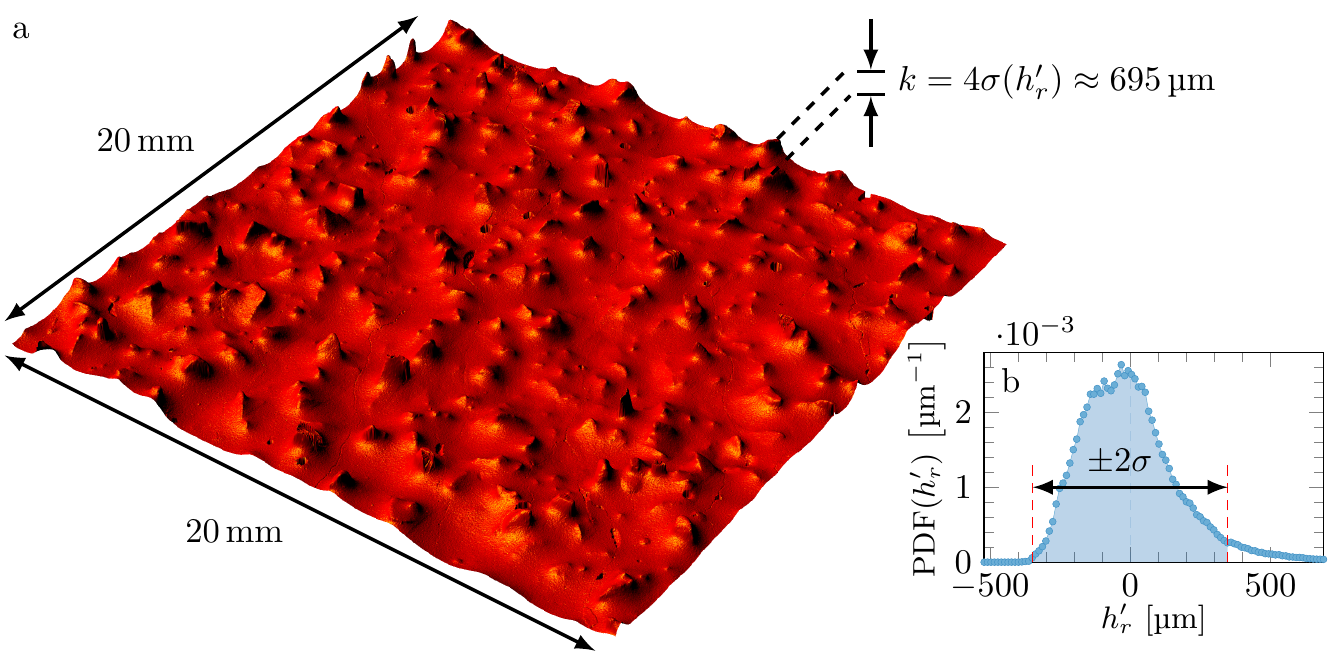}
    \caption{a) Height scan captured using confocal microscopy of a patch of sandpaper of \SI[product-units = repeat]{20 x 20}{\milli\metre} with a resolution of \SI{2.5}{\micro\meter}. The typical size of the grains is given by $k \equiv 4\sigma(h_r) = \SI{695}{\micro \meter}$ where $h_r$ is the height and $\sigma$ the standard deviation. The normalized typical grain size is then $k/d\approx 0.01$. b) Probability density function (PDF) of the measured height of the roughness stripe, with subtracted mean $h_r' = h_r - \langle h_r \rangle$.}
    \label{fig:roughness}
\end{figure}

The experiments have been performed in the Twente Turbulent Taylor--Couette (T$^3$C) facility as shown in \reff{fig:setupnum}a (details of the experimental facility can be found in~\citet{vanGils2011}). The inner cylinder has a radius $r_i~=~\SI{200}{\mm}$ and the outer cylinder has a radius $r_o = \SI{279.4}{\mm}$, such that the gap size is $d=r_o-r_i=\SI{79.4}{\mm}$, and the radius ratio {$\eta=0.716$}. The length of the cylinders is $L = \SI{927}{\milli \metre}$, which leads to an aspect ratio $\Gamma=L/d=11.7$. The outer cylinder (OC), is made of transparent acrylic which allows for optical access to the flow. The working fluid is demineralised water. We apply axially varying roughness to the inner cylinder (IC), which leads to patterns of uniformly rough and hydrodynamically smooth bands in the spanwise direction (see \reff{fig:setupnum}a). The rough stripes are made of P36 ceramic industrial grade sandpaper and are fixed to the IC using double-sided adhesive tape. In \reff{fig:roughness}, we show the height scan of a roughness element using confocal microscopy. The scan revealed that the height ($h_r$) of the roughness is mostly within $\pm2\sigma(h_r)$ of the mean, giving a characteristic length scale $k \equiv 4\sigma(h_r) = \SI{695}{\mu m}$ (see \reff{fig:roughness}b). More statistics of the roughness is shown in table\,\ref{tbl:roughness}. We fix the surface coverage of the roughness at $56\%$ such that $0.56 A_i$ of the cylinder is rough, where $A_i=2\pi r_i L$ is the area of the entire IC. The torque is only measured in the middle section {of the IC, for which the roughness surface coverage is also 56\% (see \reff{fig:setupnum}a}).
\begin{table}
\centering
\begin{tabular}{cc}
Metric & Value \\
\hline
$\sigma(h_r) = \sqrt{\left \langle h_r'^2 \right \rangle}$ & \SI{174}{\micro\meter} \\
$\left \langle \left| h_r' \right| \right \rangle$ & \SI{134}{\micro\meter} \\
$\min(h_r')$ & \SI{-527}{\micro\meter} \\
$\max(h_r')$ & \SI{738}{\micro\meter} \\
$\mathrm{median}(h_r')$ & \SI{-19.6}{\micro\meter} \\
$\mathrm{mode}(h_r')$ & \SI{-27}{\micro\meter} \\
$\mathrm{IQR}=Q3-Q1=\mathrm{CDF}^{-1}(0.75)-\mathrm{CDF}^{-1}(0.25)$ & \SI{215}{\micro\meter} \\
$\left \langle h_r'^3 \right \rangle / \left \langle h_r'^2 \right \rangle^{3/2}$ & 0.928 \\
$\left \langle h_r'^4 \right \rangle / \left \langle h_r'^2 \right \rangle^2$ & 4.361 \\
wetted area / flat area & $\approx 1.6$ \\
$h_\text{smooth}$ & \SI{-1837}{\micro\metre}
\end{tabular}
\caption{Various statistics of the roughness $h_r' = h_r - \langle h_r \rangle$ based on the data obtained from confocal microscopy, see also \protect\reff{fig:roughness}. $h_\text{smooth}$ is the distance to the smooth cylinder surface, relative to $\left \langle h_r \right \rangle$. These values represent the actual roughness used in experiments; in DNS a scaled version of these values are used.}
\label{tbl:roughness}
\end{table}%
%
\subsubsection{Global measurements: Torque}
We measured the torque $\mathcal{T}$ needed to drive the inner cylinder at constant angular velocity (the outer cylinder is kept at rest). For this we used a hollow flange reaction torque transducer connecting the drive shaft and the inner cylinder. We continuously measured the torque while quasi-statically ramping the frequency of the inner cylinder, $f_i$, from $\SI{5}{\hertz}$ to $\SI{18}{\hertz}$. This corresponds to $\Ta \approx 4\times 10^{11}$ and $\Ta \approx 6 \times 10^{12}$ {for the case of water as the working fluid}. All the experiments were performed at \SI[separate-uncertainty = true,multi-part-units=single]{21(1)}{\celsius} and {all fluid flow properties are calculated at their actual temperature}. Table\,\ref{sim_param} shows additional experimental parameters.
%
\subsubsection{Local measurements: LDA and PIV}
We performed an axial scan of the azimuthal velocity with laser Doppler anemometry (LDA). The scan was performed at the middle of the gap, $\tilde{r}=(r-r_i)/d=0.5$, at a fixed $\Ta=9.5\times 10^{11}$.
The flow was seeded using $\SI{5}{\mu \meter}$-diameter polyamide particles with {a} density of \SI{1030}{\kilo\gram\per\metre\cubed} that act as tracers \citep{vanGils2012}.
The laser beam{s} went through the outer cylinder and {were} focused in the middle of the gap. We corrected for curvature effects by numerically ray tracing the LDA beams as was shown in \citet{Huisman2012}. The axial extent of the LDA scans was $0 \leq z/L \leq 0.5$.
Particle image velocimetry (PIV) measurements were performed at $\Ta=9.5\times 10^{11}$ (same as LDA) in the radial-azimuthal plane.
The scan was performed for various heights and for all $\tilde{s}$.
For the PIV measurements the flow was seeded with different particles, namely,
{fluorescent polymer particles} (\texttt{Dantec FPP-RhB-10}) with diameters of 1--\SI{20}{\micro\metre} with a seeding density of $\approx 0.01 \ \text{particles}/\text{pixel}$.
These particles have an emission peak {with a wavelength} of $\approx\SI{565}{\nm}$. We illuminated the particles with a \texttt{Quantel Evergreen 145} \SI{532}{\nano\meter}, double pulsed laser. A cylindrical lens was used to create a light sheet of $\approx \SI{1}{\mm}$ thickness. The images were captured with an \texttt{Imager SCMOS ($2560 \times  2160$ pixel) 16 bit} camera with a \texttt{Carl Zeiss 2.0/100} lens. The camera was operated in double frame mode with a frame rate $f$ which was {much} smaller than the {inverse} interframe time $1/ \Delta t$, i.e. $\Delta t \ll 1/f$. In order to enhance the particle contrast in the images, we added a{n} \texttt{Edmund High-Performance Longpass 550 nm} filter to the camera lens. For every $\tilde{s}$, the axial extent of the experiments was different. This is done because---as will be shown later---the aspect ratio of the rolls change depending on $\tilde{s}$. For the smallest $\tilde{s}=0.63$ however, the axial resolution was $\delta z/L \approx 0.011$ while for the largest value $\tilde{s}=3.74$, $\delta z / L\approx 0.022$. Since we scan in the axial direction, the focus of the camera was changed accordingly. The fields were resolved with a commercial PIV software (\texttt{Davis 8.0}) based on a multi-step method. The initial window size was set to $64\times 64 \ \text{pixels} $ and it decreased to $32\times 32$ pixels for the last iteration. The fields are calculated in {C}artesian coordinates, which we transformed to polar coordinates. The final result were the fields in the form $\vec{u}=u_r(r,\theta,t)\hat{e_r}+u_\theta(r,\theta,t)\hat{e_\theta}$, where $u_r$ and $u_\theta$ are the radial and azimuthal velocity component which depend on the radius $r$, the azimuthal (streamwise) direction $\theta${,} and time $t$.
%
\subsection{Numerical methods}\label{sec:num_methods}
The Navier-Stokes (NS) equations were spatially discretized by using a central second-order finite-difference scheme and solved in cylindrical coordinates by a semi-implicit procedure \citep{Verzicco1996, vanderPoel2015}. The staggered grid is homogeneous in both the spanwise and streamwise directions (the axial and azimuthal directions, respectively). The wall-normal grid consists of a modified double cosine (Chebychev-type) mesh distribution. Below the maximum roughness height, we employed a cosine stretching such that the maximum grid spacing was always smaller than $0.5$ times the viscous length scale $\delta_\nu=\nu/u_\tau$. In the bulk of the fluid, we employed a second stretching, such that the maximum radial grid spacing in the bulk is  $\approx 1.7\delta_\nu$. The smallest radial mesh was $\approx 0.33 \delta_\nu$, and is located at the position of the maximum roughness height, where we expect the highest shear stress. In table\,\ref{sim_param}, we show a summary of the relevant run parameters. Time advancement was performed by using a fractional-step third-order Runge--Kutta scheme in combination with a Crank--Nicolson scheme for the implicit terms. The Courant--Friedrichs--Lewy (CFL) $(u\Delta t)/ (\Delta x)<0.8$ time-step constraint for the non-linear terms was enforced to ensure stability.
We scale the roughness stripe such that the maximum roughness height, and thus the maximum blockage ratio, was max($h_r$)$=0.1d$. Depending on $\tilde{s}$, we cut out a portion of roughness from the scanned surface. The roughness was then mirrored and {concatenated} to obtain a {streamwise homogeneous spanwise periodic stripe}. The streamwise and spanwise lengths of the computational domain are set to match the minimum computational domain size as studied in \cite{Ostilla-Monico2014b}.
A moving average over $10\times 10$ points is employed to smooth the scan from measurement noise. Finally, we set the resolution based on the demands ($\Delta z^+, r_i^+ \Delta \theta < 3$), which is small enough to recover the smallest geometrical features of the surface. 
The sandpaper roughness was implemented in the code by an immersed boundary method (IBM) \citep{Fadlun2000}. In the IBM, the boundary conditions were enforced by adding a body force $\mathbf{f}$ to the Navier-Stokes equations. A regular, non-body fitting, mesh can thus be used, even though the rough boundary has a very complex geometry. We perform interpolation in the spatial direction preferential to the normal surface vector to transfer the boundary conditions to the momentum equations. The IBM has been validated previously \citep{Fadlun2000, Iaccarino2003, Stringano2006, Zhu2016, Zhu2017, Zhu2018}. 
\begin{table}
\centering
\footnotesize
\hspace*{-0.0cm}\begin{tabular}{ccccc|ccc|cccccccccc}
$\tilde s$ & $N_\theta\times N_z \times N_r$ &$\Ta$ & $\Gamma$  & $\text{Re}_\tau$ & $C_f$ & $\Nuw$ & $\Delta (\omega)^+$ &  $\Delta r^+_\text{min}$ & $\Delta r^+_\text{max}$ & $t_{av}/T$ \\
\hline
Simulations &  & $\times 10^9$ \\ 
\hline
Smooth 	 & $758\times600\times840$  & $2.39$  & $2.08$& $697$  & $0.049$  &$30.1$  & $-$  & $0.28$  & $2.44$ & {$100$}\\
$0.47$   & $1324\times 1275\times 1200$ & $1.33$ & $2.52$ & $686$ &$0.084$  & $38.9$ & $6.36$  & $0.33$  & $1.76$ & {$43$}\\
$0.61$   & $1324\times 1121\times 1200$ & $1.33$ & $2.22$ & $690$ &$0.085$  & $39.4$ & $6.78$  & $0.33$  & $1.77$& {$42$} \\
$0.93$   & $1324\times 1682\times 1200$ & $1.45$ & $3.32$ & $692$ &$0.079$  & $38.1$ & $6.21$  & $0.33$  & $1.77$& {$42$} \\
$1.23$   & $1324\times 1121\times 1200$ & $1.37$ & $2.22$ & $685$ &$0.082$  & $38.3$ & $6.20$  & $0.33$  & $1.75$& {$30$} \\
Rough& $1324\times 1012\times 1200$ & $1.19$ & $2.00$ & $689$ &$0.095$  & $41.5$ & $8.11$  & $0.33$  & $1.76$& {$43$} \\
\hline
Experiments &  & $\times 10^{12}$ & & $\times 10^3$\\ 
\hline
Smooth & & 1.00 & 11.7 & 10.1 & 0.024 & 307 & & & \\
0.61 & &$1.00$ & $11.7$   & $13.0$  & $0.041$ & $509$ & &  &\\
0.93 & &$1.00$ & $11.7$   & $13.3$  & $0.043$ & $534$ & &  &\\
1.23 & &$1.00$ & $11.7$   & $13.0$  & $0.041$ & $508$ & &  &\\
1.87 & &$1.00$ & $11.7$   & $12.8$  & $0.039$ & $491$ & &  &\\
3.74 & &$1.00$ & $11.7$   & $13.1$  & $0.041$ & $517$ & &  &\\
Rough & &$1.00$ & $11.7$   & $13.9$  & $0.046$ & $578$ & &  &\\
\hline
\multicolumn{8}{c}{} \\
\end{tabular}
\caption{List of parameters involved in both the simulations and the experiments.
$\tilde{s}=s/d$ is the normalized roughness width with smooth and rough indicating the fully smooth and fully rough case respectively.
$N_\theta\times N_z \times N_r$ is the numerical resolution in the azimuthal, axial, and radial direction, respectively. {$\Gamma=L/d$ the aspect ratio and $L_\theta=r_i\frac{1}{3}\pi$ is the constant azimuthal length of the domain.} $\Delta (\omega)^+$ is the downward shift of the angular velocity profile $\omega^+$. $\Delta r_{min}^+$ is the minimum spacing in the wall normal direction at the location of the maximum roughness height. $\Delta r_{max}^+$ is the maximum spacing in the wall normal direction. $r_i^+ \Delta \theta = \Delta z^+ \approx 2.7$ ($r_o^+ \Delta \theta \approx 3.8$) is the grid spacing in the streamwise and spanwise directions. In the DNS, the roughness height $k^+ = 4\sigma(h_r)^+ = 130 \pm 1$ for all rough cases. {$t_{av}/{T}$ is the averaging time needed to collect statistics, normalized with the bulk flow time scale $T=d/(r_i(\omega_i-\omega_o))$.}}
\label{sim_param}
\end{table}

\section{Results}\label{sec:spanwiseresults}
\subsection{Response of the Turbulent Taylor Vortices}\label{sec:resultsec}
\begin{figure}
\centering
\includegraphics{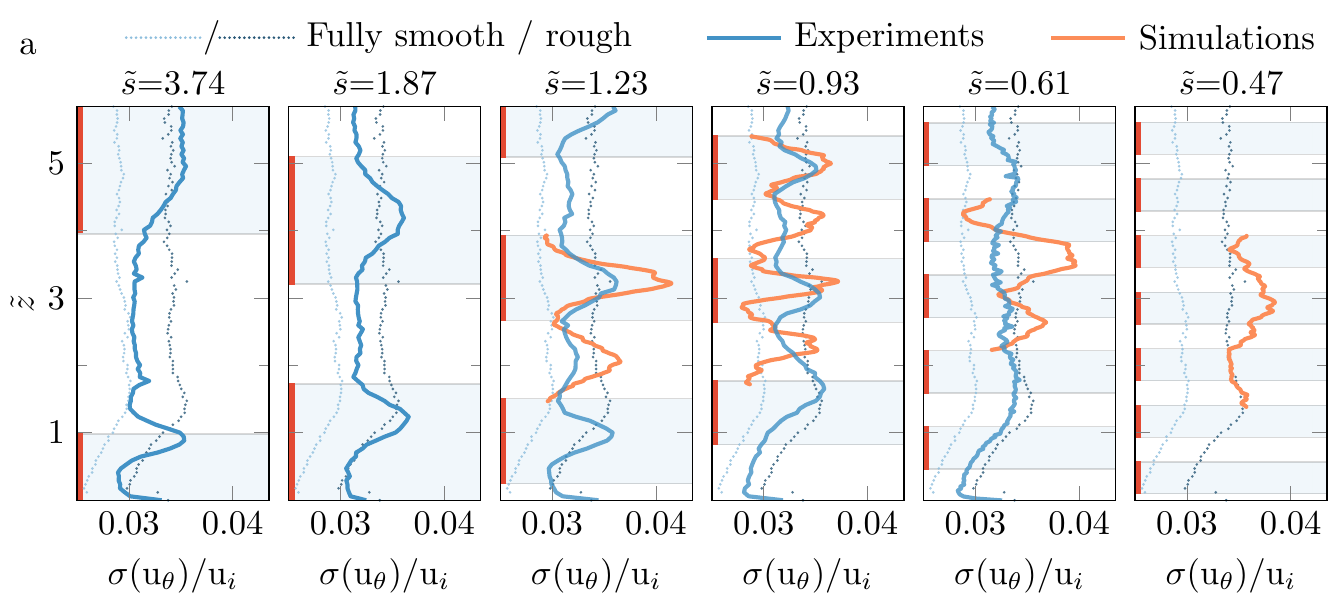}
\includegraphics{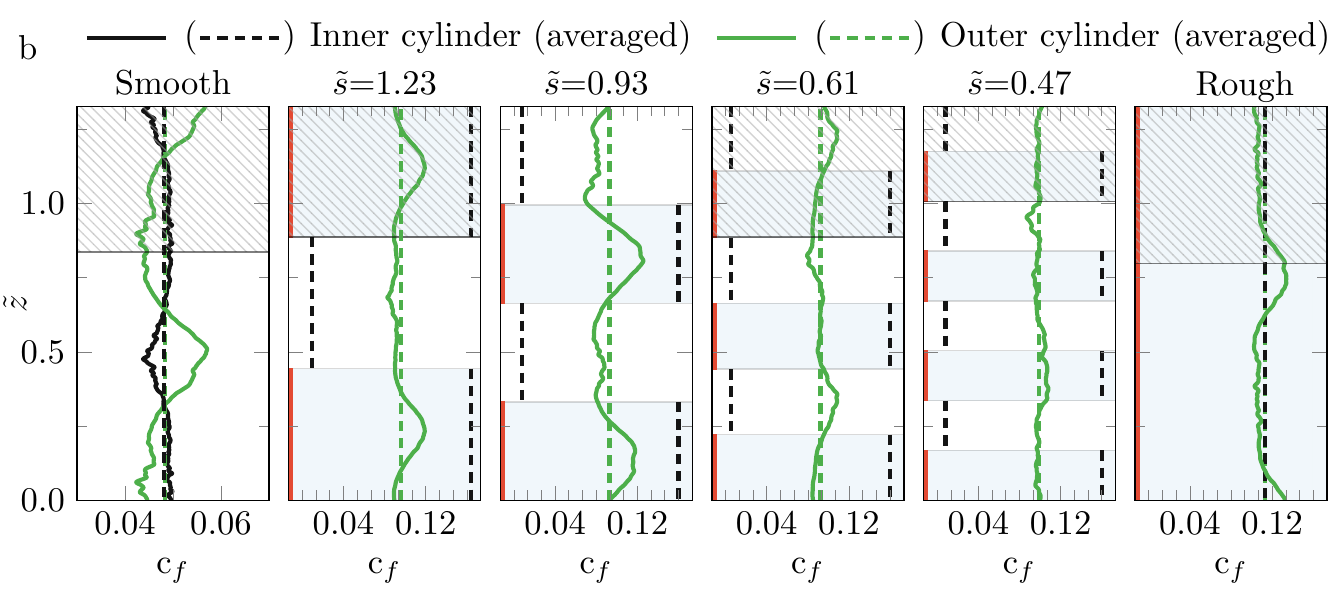}
\caption{(a) {Normalized} standard deviation of the azimuthal velocity $\sigma(u_\theta)$/$u_i$ at mid gap, as a function of $\tilde {z}=z/d$ for various $\tilde{s}$. $\Ta=\num{1e12}$ for all experiments {and $\Ta=\mathcal{O}(10^9)$ for the DNSs}. The enforced roughness pattern is indicated in red and a light blue shade. The signature of the roughness pattern is clearly visible in the bulk flow, {both for the numerical simulations (orange) as for the experiments (blue)}. For $\tilde s=0.61$, the roughness pattern does not leave a distinct imprint of its topology in the midgap flow statistics.
{(b) Local friction factor $c_f(\tilde{z})$ versus the axial height $\tilde{z}=z/d$ for $\Ta=\mathcal{O}(10^9)$.
The black lines show $c_f(\tilde{z})$ at the IC and the green line show $c_f(\tilde{z})$ at the OC.
$c_{f,r}$ above the rough patches was calculated by subtracting the smooth average of $c_{f,s}$ from $C_f=\langle c_f(\tilde{z})\rangle_L$ of the entire IC.
Hatched regions indicate axially translated copies of the same data (including averages) -- possible due to the periodic boundary condition in the axial direction -- to allow for straightforward comparison.
Imprints of the large secondary flows on the friction at the cylinder walls is observed, where impacting region experience a higher shear stress.
}}
\label{fig:explda}
\end{figure}
In order to get a first insight on the effect of the roughness on the flow, we performed axial scans of the azimuthal flow velocity at midgap using LDA. Subsequently, we calculated the standard deviation of the azimuthal velocity. In \reff{fig:explda}, we show the standard deviation of the azimuthal velocity $\sigma(u_\theta)$ normalized by the inner cylinder velocity $u_i$, as a function of the height, for various $\tilde s$.
Here, the axial coordinate is normalized using the {cylinder gap width $d$} such that $\tilde z = z/d$.
\refF{fig:explda} reveals that for the case of the largest stripe {width} ($\tilde s=3.74$), the smooth section has, on average, a value of $\sigma(u_\theta)/u_i  \approx 0.03$, slightly larger than
we found for the fully smooth case (shown with the dotted light blue lines).
Above the rough {stripe}, towards the center of the setup (\text{i.e.} for large $\tilde z$), $\sigma(u_\theta)$ gradually increases to a value of approximately $\sigma(u_\theta)/u_i \approx 0.04$. A similar, but not so clear trend can be seen at the lower roughness section ($\tilde z \approx 0.93$) of this case. However, this might be influenced by the lower bottom plate of the system.

When looking at the $\tilde s=1.87$ case, we see very similar, however more pronounced dynamics. Azimuthal velocity fluctuations are promoted in regions where the roughness is present, as suggested by the appearance of local peaks centered at the position of the rough stripes. This effect is further seen for the cases of $\tilde{s}=1.23$ and $\tilde{s}=0.93$, where we observe similar profiles. At their smooth areas however, {we also observe enhanced velocity fluctuations, albeit less pronounced than the locations above the rough patches. This effect is not seen for $\tilde{s}>1.27$}. For the final case with $\tilde s=0.61$ these trends seems to fade away and we see that $\sigma(u_\theta)$ becomes more axially independent, i.e. the peaks are less pronounced, and do not seem to follow the topology of the roughness stripes.

\begin{figure}
\centering
\includegraphics[width=0.99\textwidth]{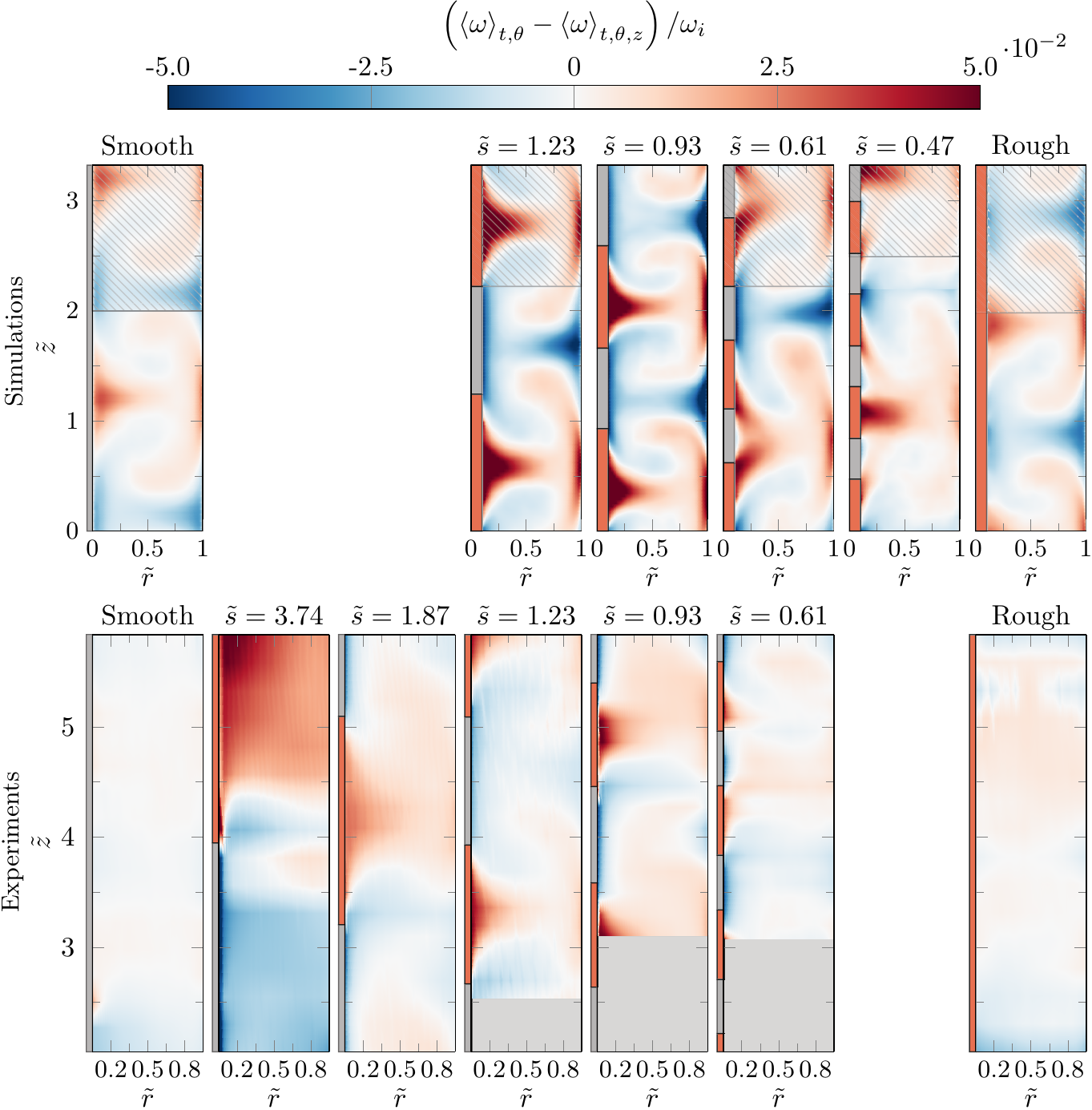}
\caption{%
Deviation of the temporal and azimuthally averaged angular velocity $\langle \omega \rangle_{t,\theta}$ with respect to the temporal, azimuthal, and axial averaged angular velocity $\langle \omega \rangle_{t,\theta,z}$ obtained from DNS at $\Ta\approx \mathcal{O}(10^{9})$ (top), and experiments at $\Ta=\num{1e12}$ (bottom), for various $\tilde s$ explored.
For experiments, $\tilde r$ spans between $\left[ 0.05, 0.95 \right]$.
All fields are normalized with the angular velocity {of the inner cylinder} $\omega_i=u_i/r_i$.
Positive values represent velocities that are closer to the IC velocity.
The leftmost panel corresponds to the case of no roughness (smooth) while the rightmost panel is the case where the entire IC is uniformly rough.
For better comparison, overlapping $\tilde s$ cases for DNS and experiments are aligned vertically.
Missing cases are were not feasible in experiments or DNS.
{Hatched regions in the DNS figures indicate axially translated copies of the same data---possible due to the periodic boundary condition in the axial direction---to allow for straightforward comparison.}
The gray shaded areas in the gap represent unexplored heights.
Ejecting regions can be seen in axial locations where the roughness is present.
Notice the similarity of the {flow} structures between DNS and experiments.
}
\label{fig:simomega}
\end{figure}

{The results of the DNSs, presented together with the experiments in \reff{fig:explda} exhibit very similar behaviour. When normalized with $u_i$, we find that the standard deviation of the streamwise velocity show similar values as in the experiments. This is intriguing since the $k/d$ values in the simulations are almost one order of magnitude larger than in the experiments. Above the rough {stripes}, we find enhanced $\sigma(u_\theta)/u_i$, whereas over smooth {stripes} we find diminished $\sigma(u_\theta)/u_i$. For $\tilde{s}=0.61$ however, the trends are somewhat different. There, we find enhanced $\sigma(u_\theta)/u_i$ above the smooth region in the DNSs, which is explained by the recombination of plume ejection regions, forming one larger TTV above the smooth regions, see \reff{fig:simomega}.}

{The findings presented in \reff{fig:explda} show that the presence of the roughness affects the relative turbulence statistics in the bulk of the flow, far away from the roughness sublayer region} {(in contrast to homogeneous roughness in TC flow} \citep{Berghout2018}), reminiscent to what is found in studies of pipe and channel flow \citep{Koeltzsch2002,Chung2018}.

{\refF{fig:explda}(b) shows the axial variations of the friction factor $c_f(\tilde{z})$ (see \refsec{sec:spanwiseintro}) on both the inner cylinder and the outer cylinder. The solid and dashed black lines represent $c_f(\tilde{z})$ measured on the inner cylinder, and the solid (green) lines represent $c_f(\tilde{z})$ on the outer cylinder. We average both in time and in the azimuthal direction. Significant variations in $c_f(\tilde{z})$ are observed which are linked to the orientation of the TTV.
For impacting regions, \textit{i.e.}\ plumes impacting on the inner cylinder boundary layer, the wall shear stress in the azimuthal direction is enhanced.
When plumes eject from the inner cylinder boundary layer, also known as the ejecting regions, the shear stress is reduced. This, once more, illustrates the relative strength of the secondary flow (TTV) and the mean flow. For $\tilde{s}=0.93$, the variations of the friction factor on the outer cylinder are even more pronounced, thus indicating that the strength of the TTV are enhanced with enforcing the spanwise variations in the roughness. For $\tilde{s}=0.47$, the variations are not visible and the TTV are severely weakened, but still present, see \reff{fig:simomega}. }

\begin{figure}
\centering
\includegraphics[width=0.99\textwidth]{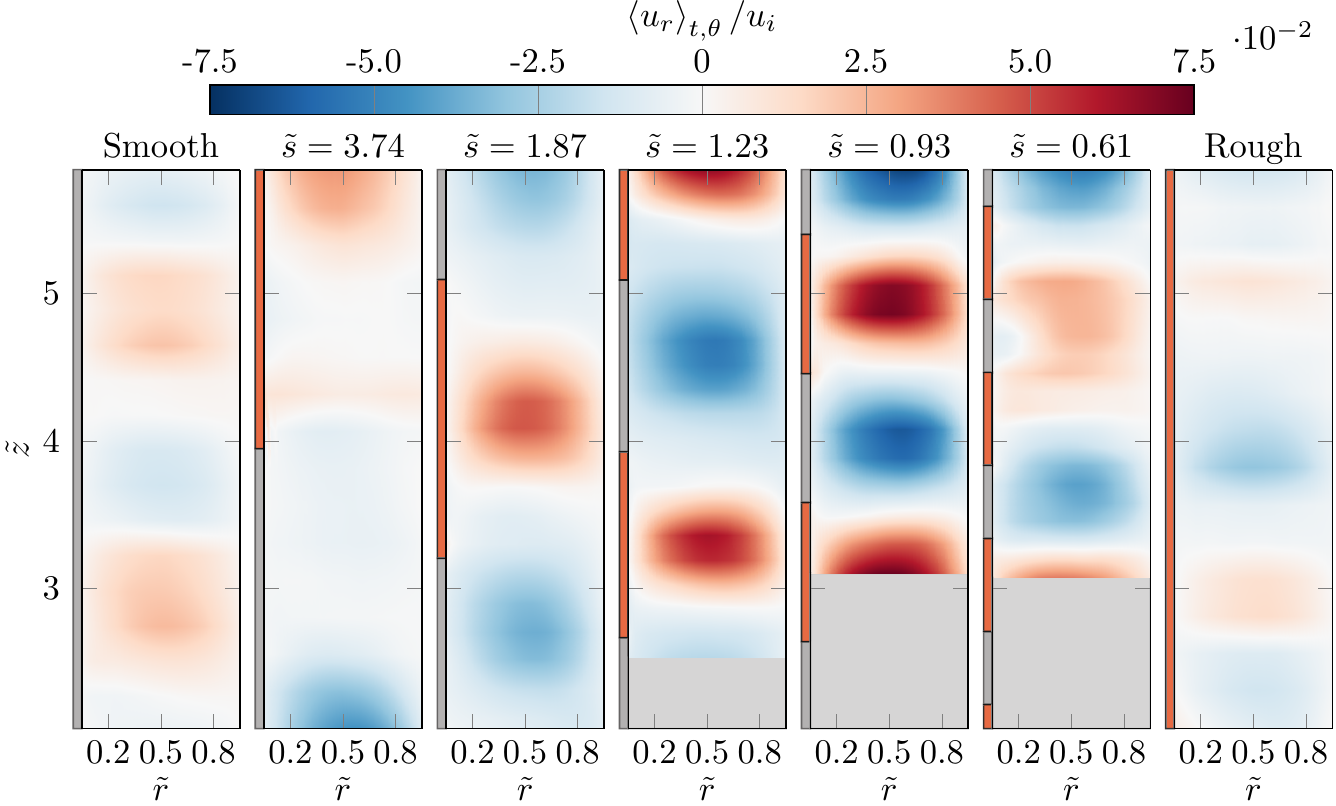}
\caption{Temporal and azimuthal average of the radial velocity $u_r$, normalized with the inner cylinder azimuthal velocity $u_i$, obtained from {experiments at $\Ta=1\times 10^{12}$ using} PIV for varying roughness stripe sizes $\tilde{s}$. A positive value of $u_r$ denotes outflow, while a negative value denotes inflow, with respect to the inner cylinder. It can be seen that the rolls are pinned by the roughness and their wavelength changes with $\tilde{s}$.
The red and gray areas at the left side of each plot indicate the positions of the rough and smooth areas, respectively.
Note that the typical grain size is $k/d \approx 0.01$. The gray shaded areas in the gap represent unexplored heights.}
\label{fig:exppiv}
\end{figure}
To gain more insight into how the roughness alters the flow, we set out to measure the velocity field in the meridional plane using PIV at multiple heights. In \reff{fig:exppiv}, we show the temporal and azimuthally averaged radial velocity component $u_r$, normalized with $u_i$, in the spanwise wall-normal plane ($\tilde z$-$\tilde r$), where the radial coordinate is normalized such that $\tilde r = (r - r_i) / d$. \refF{fig:exppiv} shows that for the case of $\tilde{s}=3.74$, a very large structure can be seen, which consists of a large outflow region (positive $u_r$) around $\tilde{z}=5.84$, while a large inflow region (negative $u_r$) is detected around $\tilde{z}=2.10$. The situation is more pronounced for the cases of $\tilde{s}=1.87$, $\tilde{s}=1.23$, and $\tilde{s}=0.93$, where a clear roll-like  structure (i.e. the TTV) can be observed. Note that the radial component in the flow changes sign along the axial direction as it should in the presence of a TTV. What is rather remarkable is that the wavelength of the rolls $\lambda$ changes for different values of $\tilde{s}$. For the large structure at $\tilde{s}=3.74$, the normalized wavelength is $\tilde{\lambda}=\lambda/d \approx 4.01$. As $\tilde{s}$ decreases to $\tilde{s}=1.87$, $\tilde{\lambda}\approx 1.49$.  At $\tilde{s}=1.23$, the wavelength decreases to a value of $\tilde{\lambda}\approx 1.42$. At $\tilde{s}=0.93$, $\tilde{\lambda}\approx 0.94$, and finally for the smallest value of  $s=0.61$, the wavelength increases slightly to $\tilde{\lambda}=1.09$. We remind the reader the work of \citet{Huisman2014}, {who} revealed that for counter-rotation {($-\omega_o/\omega_i \approx 0.4$)}, the average wavelength of the rolls could be either $\tilde{\lambda}=1.46$ or $\tilde{\lambda}=1.96$ depending on the \textit{state} attained by the system. The current work shows that by an appropriate choice of $\tilde{s}$, the wavelength of the rolls can firstly, abandon its natural state; and secondly, it can be tuned within the range $\tilde{\lambda}\in[0.94,4.01]$. The wavelengths described above were calculated by measuring the locations of two consecutive maxima and minima of $\langle u_r \rangle_{t,\theta,r_{bulk}}$ along $z$ which are closest to midheight. Here, the symbol $\langle \cdot \rangle_{t,\theta,r_{bulk}}$ denotes average over time, the streamwise direction and the bulk region, i.e. $(r_{bulk}-r_i)/d\in[0.3,0.7]$.

In addition, we observe that outflow regions (flow away from the IC) are created in axial regions where the roughness is located; and conversely, inflow regions (flow towards the IC) are created in the smooth areas. Note that this  orientation of the secondary flows is opposite to what is found in other canonical systems (e.g. pipe flow and channel flow \citep{Willingham2014, Yang2017, Chung2018}), where one finds inflow regions above the rough stripes and outflow region above the smooth stripes. {%
Consistent with findings in boundary layers however, is the correlation between low momentum pathways (LMPs) and outflow regions \citep{vanderWel2015}. Indeed we find that LMPs - with respect to the IC and located on top of the roughness - is associated with outflow regions, similar to \citeauthor{vanderWel2015} although their Lego\textsuperscript{\textregistered} roughness strips are much thinner and protrudes farther into the flow.
}

Another interesting observation is that since the driving is now from the BL rather than the bulk, the strength of the rolls changes depending on the value of $\tilde{s}$, as evidenced by the magnitude of $|u_r|$. In order to explore this feature in more detail, we quantify the strength of the rolls {by} $\tilde{u_r}^\prime\equiv \sqrt{\langle (u_r/u_i)^2 \rangle_{t,\theta,r_{bulk},z_\lambda}}$ as a function of $\tilde{s}$. Here, the symbol $\langle \cdot \rangle_{t,\theta,r_{bulk},z_\lambda}$ denotes an average over time, the streamwise direction, the bulk region, and the axial region that defines the wavelength of a single roll $z_\lambda$. In \reff{fig:exptok}(c), we show $\tilde{u_r}^\prime$ as a function of $\tilde{s}$, where we observe that the strength of the rolls increases with decreasing $\tilde{s}$ for $\tilde{s}\in[0.93,3.74]$. However at $\tilde{s}=0.61$ the trend is broken, and we observe that $\tilde{u_r}^\prime$ decreases with respect to the case of $\tilde{s}=0.93$.

In order to obtain more insight into the mechanism(s) that lead to the variation of $\tilde{\lambda}$ with $\tilde{s}$, we turn to DNS, albeit at a much lower Ta ($\approx 1.0\times 10^9$), and much higher roughness height ($k/d \approx 0.1$). Since very large $\tilde s$ cases are not feasible for DNS, we focus on matching the exact $\tilde s$ in the lower range. We will show that, despite the $\mathcal{O}(10^3)$ difference in $\Ta$, the same observations found in the numerics are also found in the experiments. 

First, we look at the azimuthal velocity component. In \reff{fig:simomega}, we plot the difference of the temporal and azimuthal average of the angular velocity {$\langle \omega \rangle_{t,\theta}$} with respect to the temporal, azimuthal and,  axial average of the angular velocity {$\langle \omega \rangle_{t,\theta,z}$}.
This is done to emphasize the underlying organization of the TTVs. Here, we clearly observe that for all $\tilde{s}$, ejecting regions of angular velocity are originated in the rough stripes, similar to the preferential plume ejection sides at the tips of grooves in \citet{Zhu2016}. These ejecting regions advect fluid from the roughness stripe on or at the inner cylinder towards the outer cylinder.
These ejecting regions occur at each roughness stripe and as a consequence, an array of plume-like structures are formed along the axial direction.
In TC flow (without roughness), plume-like structures are clear signatures of the presence of TTVs \citep{Ostilla-Monico2014b,Ostilla-Monico2014}. A closer inspection of figure \ref{fig:simomega} reveals that for the largest value of $\tilde{s}=1.23$, the plumes have enough separation not to interact among them. When $\tilde{s}$ is decreased to $\tilde{s}=0.93$, we observe that the plumes come closer, and can, in fact, interact with each other. At the lower $\tilde{s}=0.61$ however, the situation is rather different.
%
{The rough patches are too close to each other to create individual ejecting regions and therefore, merge to a larger collective plume. For the $\tilde{s}=0.61$ case, we observe in both, experiments and DNS that two ejecting regions, each located on top of a rough region, merge into a combined ejecting region. When $\tilde s$ is decreased to a value of 0.47, even three ejecting regions are combined to a single large plume. These combined plumes drive TTV with a larger wave length, therefore, decreasing the effective momentum transfer.}
These observations help us to rationalize the change in the wavelength and strength of the rolls shown in \reff{fig:exppiv}. If $\tilde{s}$ decreases, the plumes are effectively forced to come closer to each other; and as a result, the roll changes its wavelength and becomes stronger due to the added interaction of the plumes. In \reff{fig:simomega} we show $(\langle \omega \rangle_{t,\theta}-\langle \omega \rangle _{t,\theta,z})/\omega_i$, the same quantity discussed previously, albeit now for the experiments. Here, we can clearly see that a similar mechanism takes place. Plume-like structures are originated at the centers of the roughness elements and interact with each other if the spacing (small $\tilde{s}$) is reduced. 

The LDA, PIV and DNS explored in this section reveal that there is a mean effect of the spanwise-varying roughness on the large scale secondary flows that exist in turbulent TC flow. We have seen thus far that the roughness pins the rolls, and that their wavelength and strength can be tuned depending of the choice of $\tilde{s}$ over a wide range of $\Ta$, and a wide range of roughness heights $h$. However, how does the flow respond globally, i.e. the angular momentum transport, to this change in morphology? This will be addressed in the following section.

\subsection{Global response}%
\label{sec:resultglob}%
\begin{figure}
\centering
\includegraphics[width=0.49\textwidth]{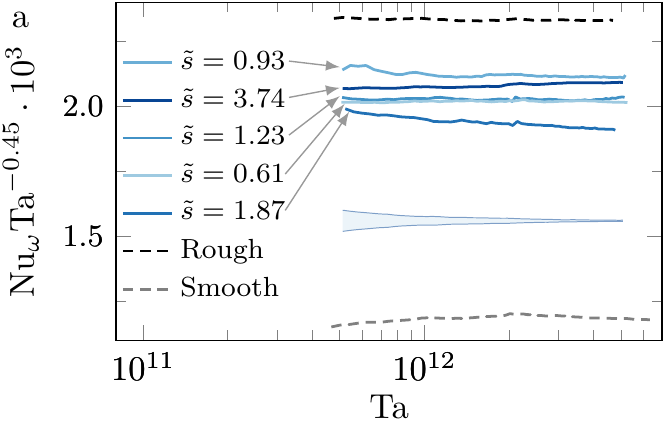}
\includegraphics[width=0.49\textwidth]{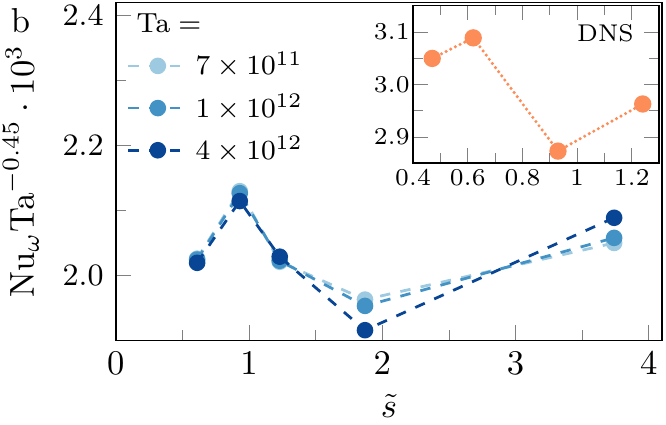}\\
\vspace{2mm}%
\includegraphics[width=0.49\textwidth]{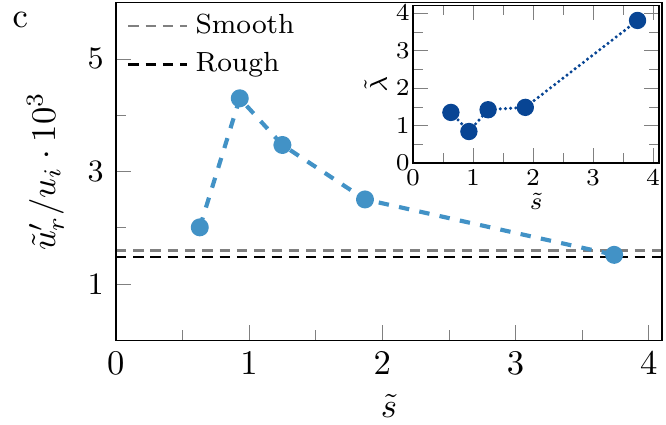}
\includegraphics[width=0.49\textwidth]{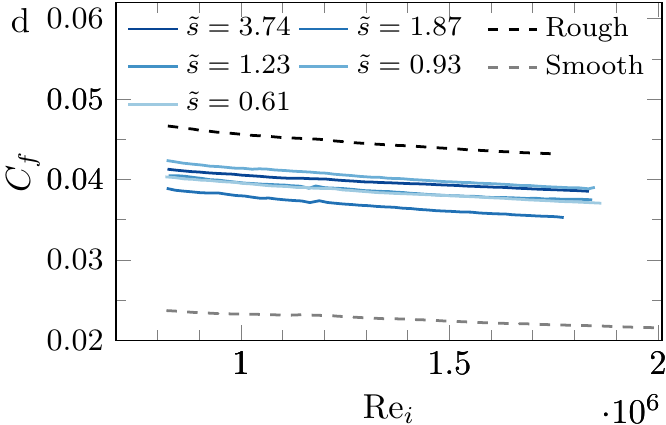}
\caption{(a) Compensated Nusselt number $\Nuw \Ta^{-0.45}$ as function of $\Ta$ for varying $\tilde{s}$. The shaded area indicates the error of the measurements, which can be seen to decrease with increasing driving {strength}. (b) Compensated Nusselt number $\Nuw \Ta^{-0.45}$ as a function of $\tilde{s}$ for {three} selected \Ta. Here, an optimum value in the transport of angular momentum is observed close to $\tilde s \approx 1$. The inset in (b) shows the results obtained with the DNS ({$\Ta\approx 10^9$}), where the maximum can be observed at a {slightly} lower $\tilde{s}$, namely $\tilde{s}\approx 0.6$. (c) The strength of the rolls, quantified as the normalized RMS of the radial velocity $\tilde u_r'$ as a function of $\tilde{s}$, obtained from the PIV experiments. The inset shows the wavelength of the TTV as function of $\tilde s$. (d) Friction coefficient $C_f$ as a function of the driving {strength}, expressed with the Reynolds number $\text{Re}_i$, for various $\tilde{s}$.}
\label{fig:exptok}
\end{figure}

The global response of the TC system can be expressed with  $\Nuw$ (\refe{eq:spanwisenu}) or with the related friction coefficient $C_f$.
In figure \ref{fig:exptok}(a), we show the compensated $\Nuw$ as a function of the driving {strength} $\Ta$, where a scaling of $\Nuw \propto \Ta^\alpha$, with $\alpha=0.45$ is observed for all the $\tilde{s}$ explored.
The same data is represented as $C_f$ versus $\Rei$ in \reff{fig:exptok}(d).
In absence of roughness and within the same range of Ta, the scaling is found to be effectively {$\Nuw\propto \Ta^{0.39}$} \citep{Paoletti2011,vanGils2011,Huisman2014}.
In contrast, when both of the solid walls are made uniformly rough (i.e. pressure drag dominates), the scaling asymptotes to the ultimate regime predicted by Kraichnan, \textit{i.e.} $\Nuw\propto \Ta^{0.5}$ \citep{Kraichnan1962,Zhu2018}. In  \cite{Zhu2018}, the closest configuration to our study is the case of rough IC and smooth OC, for which {an effective} exponent $\alpha=0.43$ was found. We note that this exponent is slightly smaller than the ones observed in the current study. The reason behind this is currently unknown. We notice, however, that the roughness type in our study is rather different. {In this study we use spanwise-varying sand grain roughness, while the roughness in \citet{Zhu2018} is made of rib obstacles and is oriented perpendicular to the streamwise direction.}

In order to connect the observed dynamics of the TTVs with the global response, we plot in \reff{fig:exptok}(b) the compensated Nusselt number $\Nuw \Ta^{-0.45}$ as a function of $\tilde{s}$ for both the experiments and the numerics. We note that the exponent found for $\tilde{s}=1.87$ ($\alpha=0.44$) is nearly the same as $\alpha=0.45$. We notice rather remarkably, the appearance of a maximum around $\tilde{s}\approx 0.93$ for the experiments, and $\tilde{s} = 0.61$ for the DNS (shown in the inset of \reff{fig:exptok}(b)). We attribute the appearance of this peak to the strengthening of the TTVs, which is caused by the variation of $\tilde{s}$, and thus of $\tilde{\lambda}$. Explicitly, by lowering $\tilde{s}$, we can decrease the wavelength of the rolls, as seen in the inset of \reff{fig:exptok}(c), thereby, bringing them closer together (see also \refsec{sec:resultsec}). As a consequence, in contrast the rolls are strengthened which leads to an enhancement of the angular momentum transport; and thus, the peak around $\tilde{s} = 0.93$.
{This peak is also visible in the normalized RMS of the radial velocity, $\tilde u_r'$, plotted in \reff{fig:exptok}(c). With decreasing $\tilde s$, $\tilde u_r'$ increases until an optimum is been reached around $\tilde s = 0.93$. Below the optimum, $\tilde u_r'$ drops drastically to much lower values.}
This {optimum} is also observed by \citet{Huisman2014}, though the {mechanism leading to optimum transport} there is quite different. While the rolls in their study are enhanced by counter-rotating the OC; in our case, the rolls are strengthened by forcing {$\tilde{\lambda}$} below their natural wavelength due to the right choice of the {size of the} spanwise varying roughness {$\tilde s$}. This is also supported by the observation that the magnitude of the radial velocity shows a maximum around $\tilde{s}=0.93$, as shown in \reff{fig:exptok}(c). We note, however, that the torque is not measured throughout the entire axial length of the cylinders $L=\SI{927}{\mm}$, but in a smaller section of length $L_{\text{mid}}=\SI{536}{\mm}$. As a result, the large structure identified previously for the case of $\tilde{s}=3.74$ ($\tilde{\lambda}=4.01$), does not fit entirely in the measurement section (see the first panel of \reff{fig:exppiv}). As a result, the Nusselt number corresponding to this case could be under or overestimated.

We also note that in the case of the numerics, the position of the maximum is different than in the experiments. We attribute this to a combination of two effects. On the one hand the DNS is performed at a lower Ta, which has an effect on the natural wavelength of the rolls as it was shown by \citet{Chouippe2014}, who show that for similar values of $\eta$, the wavelength of the rolls can decrease with decreasing Ta. On the other hand, the axial domain of the DNS is bounded by $\Gamma\in[2.08,3.32]$, which gives rise to limited box-sizes. Thus, when $\tilde{s}$ is varied, the rolls could suffer from an additional $\textit{constraint}$ due to the limited axial domain. In addition to this discrepancy, we also note that the scaling in the range of Ta at which the DNS is done ($\approx 1.0\times 10^9$), is not known a priori.
In absence of a better choice, we compensate the numerical data using the same exponent as in the experiments (\reff{fig:exptok}(b)). However, we note that this exponent might be different due to the 2 decades of separation  in Ta between the numerics and experiments, as was also shown by \citet{Zhu2018}. We would like to emphasize, however, that in spite of these discrepancies, a maximum in angular momentum transport is observed for a given $\tilde{s}$ in both the experiments and the numerics, which is solely a consequence of the varying axial wavelength of the TTV, dictated by the spanwise-varying roughness.

\subsection{Velocity profiles}
\label{sec:results_velprofiles}

\begin{figure}
\centering
  \includegraphics[width=0.49\linewidth]{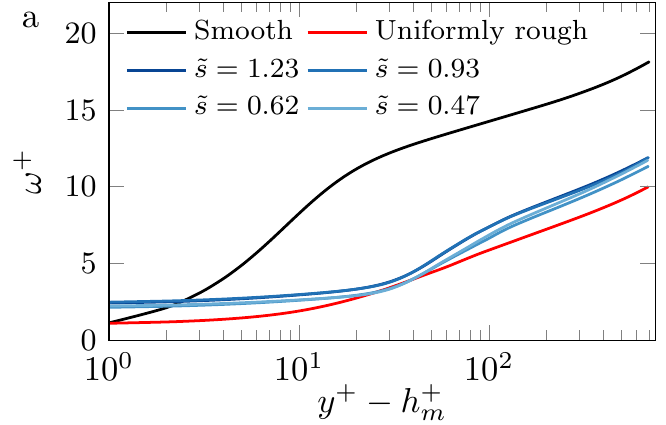}
  \includegraphics[width=0.49\linewidth]{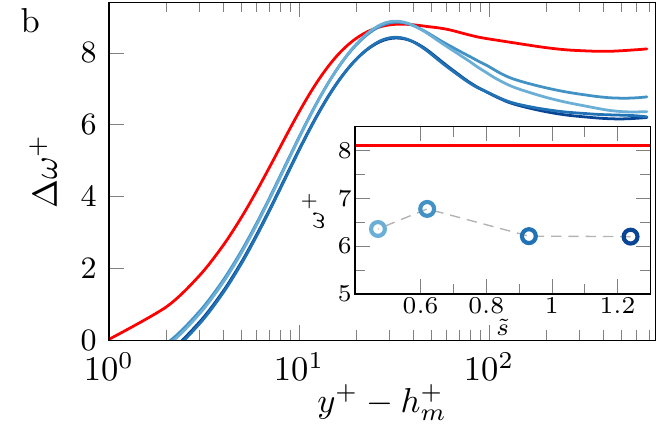}
  \vspace{3mm}%
  \includegraphics[width=0.49\linewidth]{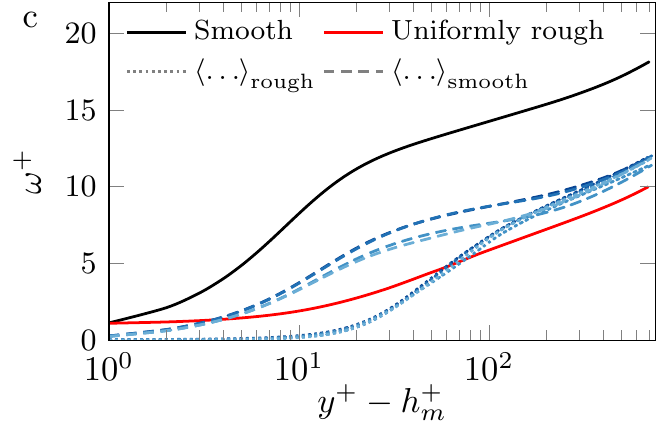}
  \includegraphics[width=0.49\linewidth]{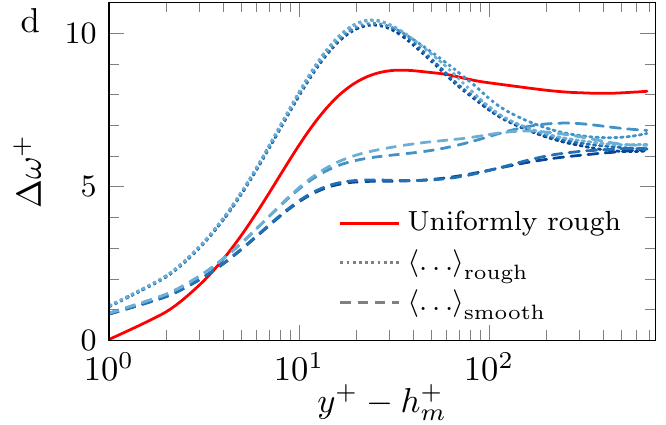}
\caption{
(a) Angular velocity $\omega^+$ profiles {in the reference frame of the IC} versus the wall normal distance $y^+-h_m^+$ for various $\tilde{s}$, where $h_m^+$ is the virtual origin and equals the melt-down (i.e. mean) height {$h_m/d\approx0.2$} of the rough surface {and $y^+ = (r-r_i)/\delta_{\nu}$}. The solid black line represents the uniformly rough case. (b) Angular velocity shift $\Delta \omega^+$ as a function of {$y^+-h^+_{m}$} for varying $\tilde s$. In the inset of (b), we show the angular velocity shift $\Delta \omega^+$ versus the wall normal distance $y^+-h_m^+$. Here, we observe a maximum downwards shift of the angular velocity profile for the simulation where we cover the entire inner cylinder with sandpaper roughness (i.e. uniformly rough). { (c) The angular velocity conditioned on the axial location: above smooth surface ($\langle \omega^+ \rangle_\text{smooth}$) and rough surface ($\langle \omega^+ \rangle_\text{rough}$). (d) The angular velocity shifts conditioned on the axial location: above smooth surface ($\langle \Delta \omega^+ \rangle_\text{smooth}$) and rough surface ($\langle \Delta \omega^+ \rangle_\text{rough}$). We conclude that the plumes originating from the roughness elements lead to enhanced mixing of streamwise momentum, and hence a downwards shift of the velocity profiles. Further away from the wall, the bulk is well-mixed and the streamwise profiles above smooth and rough wall locations converge to a similar value.}}
\label{fig:o_do}
\end{figure}

Having discussed the dynamics of the TTVs and the corresponding global response, in terms of the dimensionless torque, we now set out to study the streamwise, angular, velocity profiles (rather than the azimuthal profiles, as discussed in \cite{Grossmann2014} and \cite{Berghout2018}). To allow for straightforward comparison between the respective velocity profiles, we run the DNS at constant friction Reynolds number $Re_\tau = 690 \pm 10$. The profiles are then temporally, azimuthally, and axially averaged $\omega^+ = \langle \omega \rangle_{t,\theta,z}/\omega_\tau$. The profiles still exhibit a logarithmic region when averaged over the entire axial coordinate. \refF{fig:simomega} shows however that the TTVs in the flow, following the spanwise-varying roughness, do not exhibit any outer similarity. Deviations of the azimuthal and temporal averages from the mean logarithmic profiles are found up to $\Delta \omega^+ \approx 2$. 

For turbulent flows over rough walls, the streamwise velocity profiles retain their logarithmic form. However, the hallmark effect of rough walls is a downwards shift of this region (for any drag increasing surface), which can also be understood as an increase of the skin friction factor $C_f$ \citep{Hama1954}. \refF{fig:o_do}(a) shows the angular velocity profiles $\omega^+$ as a function of  $(y^+-h_m^+)$, where $h_m^+$ is the virtual origin and equals the melt-down (i.e. mean) height of the rough surface {and $y^+ = (r-r_i)/\delta_{\nu}$}. We choose the melt-down height of the roughness over the full inner cylinder as the virtual origin. In figure \ref{fig:o_do}(b) we show the velocity shift versus the wall normal distance. The inset gives a vertical cut at $y^+ = Re_\tau$. It is evident that also in this representation, an optimum in the velocity shift, and thus in  $C_f$ can be observed. The position of this maximum ($\tilde s = 0.61$) is the same as the one obtained from the angular momentum transport (see \refsec{sec:resultglob}).

{In figures \refF{fig:o_do}c and \refF{fig:o_do}d we employ conditional averaging of the angular velocity profiles over the smooth $\langle \dots \hspace{-0.1mm} \rangle_\text{smooth}$ and rough $\langle \dots \hspace{-0.1mm} \rangle_\text{rough}$ axial locations. The wall shear stress in the viscous normalization is taken over the entire axial height $L$ of the inner cylinder. $h_m^+$ is also taken as a local variable. We already deduce from figure \ref{fig:simomega} that significant variations in the temporal and azimuthal average of the velocity profiles are expected, at least close to the roughness. Indeed we find that for all $\tilde{s}$ the region above the roughness is better mixed, due to the presence of plume-like structures originating from the rough surface. The angular velocity profiles is thus shifted downwards in comparison to the average over the entire IC. For the smooth wall conditioned profiles, we observe the opposite, such that the profiles lay higher.}

{The merging of plumes from different rough patches into a large scale coherent TTV is also observed in the cross-over of $\langle \omega^+ \rangle_\text{smooth}$ and $\langle \omega^+ \rangle_\text{rough}$ for $\tilde{s}=0.61$ in \refF{fig:o_do}d at $y^+-h_m^+ \approx 210$. Further into the bulk flow, turbulent processes mix out the in-homogeneous effects of rough wall attached plumes, and the angular velocity profiles converge to similar values. However, we note that even at $y^+-h_m^+ = Re_\tau$, $\langle \omega^+ \rangle_\text{smooth}$ and $\langle \omega^+ \rangle_\text{rough}$  differ to $\approx 0.5$.}

\FloatBarrier
\section{Conclusions and outlook}
\label{sec:conclusions}
{In conclusion}, we have investigated, both numerically and experimentally, large Taylor number Taylor--Couette flow in the presence of spanwise-varying roughness, which consists of an arrangement of stripes of width $\tilde s$, that covers the entire circumference of the inner cylinder. In the experiments, the stripes were made from sandpaper, while in the numerics a confocal microscopy scan of the surface was implemented by means of the immersed boundary method (IBM). 

Remarkably, we have found that by varying $\tilde s$ in the range $\tilde s = [0.47, 3.74]$ we can alter the axial wavelength  of the turbulent Taylor vortices within the range $\tilde{\lambda}\in[0.94,4.01]$, even if the roughness height was very low ($k/d\approx 0.01$). This manipulation was observed to hold in a range of three decades in Ta ($\mathcal{O}(10^9)-\mathcal{O}(10^{12})$).

In the experiments, the scaling of the Nusselt number with the driving {strength} was found to be effectively $\Nuw\propto \Ta^{0.45}$ for $\Ta\in[5\times 10^{11},5\times 10^{12}]$).
The experiments and DNSs also revealed that inflow regions ($u_r<0$) originated between the rough stripes, where the inner cylinder was hydrodynamically smooth (in contrast to secondary flows induced by spanwise-varying roughness in channel flow, where the orientation of the vortices is reversed \citep{Chung2018}. Conversely, at the center of the rough stripes, we observed the creation of outflow regions ($u_r>0$) which were accompanied by the promotion of azimuthal velocity fluctuations $\sigma(u_\theta)$ at midgap. At these axial locations (center of rough stripes), we observed, in both the numerics and experiments, the emission of plume-like structures, which are responsible for the creation and pinning of the rolls. Since the coverage of the roughness was fixed, we showed that by reducing $\tilde{s}$, we can effectively bring these structures closer, and enhance the interaction of the rolls, as evidenced by the increment in $|u_r|$. As a consequence of this interaction, the flow responded globally by inducing a maximum of angular momentum transport at $\tilde{s}=0.93$ in the experiments, and $\tilde{s}=0.61$ in the numerics.  

We wish to stress that in this study the change in the morphology of the large-scale structures is only due to the spanwise-varying roughness (of very low height) and not by a change of $\Gamma$ or $\eta$, which opens the possibility of exploring different configurations in which the rolls can be tuned at such large turbulence levels. 
 
Many questions arise from the aforementioned observations. Understanding the mechanisms leading to the merging of plume ejection regions, and accompanied parameter boundaries at which this occurs, would lead to a further insight into the dynamics of the TTVs. Furthermore, it would be intriguing, in the spirit of \citet{Bakhuis2018b}, to study the influence of spanwise-varying regions of idealized high and low wall shear stress, without geometrical induced disturbances. It is an open question whether one could also alter $\lambda$, without the interaction of the plumes.

\section*{Acknowledgements}
We would like to thank Jelle Will and Dominik Krug for various stimulating discussions. We like to thank Jos\'e Encarnaci\'on Escobar for performing the confocal microscopy measurements and Gert-Wim Bruggert for technical support.
This work was funded by Natural Science Foundation of China under grant no. 91852202, VIDI grant No. 13477, STW, FOM, MCEC, and the Netherlands Organisation for Scientific Research (NWO).
This project is also partially funded by the Priority Programme SPP 1881 Turbulent Superstructures of the Deutsche Forschungsgemeinschaft. We also acknowledge PRACE for awarding us access to MareNostrum based in Spain at the Barcelona Supercomputing Center (BSC) under PRACE project number 2017174146.
This work was partly carried out on the national e-infrastructure of SURFsara, a subsidiary of SURF cooperation, the collaborative ICT organization for Dutch education and research.

\bibliographystyle{jfm}
\bibliography{JFM_PatchedRoughness}
\end{document}